\documentclass[12pt]{article}
\usepackage{epsf}
\def\beq{\begin{equation}}
\def\eeq{\end{equation}}
\def\bea{\begin{eqnarray}}
\def\eea{\end{eqnarray}}

\def\vel{\left|}
\def\ver{\right|}

\def\ga{\left(}
\def\dr{\right)}

\def\la{\langle}
\def\ra{\rangle}
\def\ba{\begin{array}}
\def\ea{\end{array}}

\def\ds{\displaystyle}

\title{ {\bf
Exclusive $B\rightarrow K^* l^+ l^-$ decay in the three  Higgs doublet model}}
\author{\vspace{1cm}\\
        {\bf E. O. Iltan}
        \thanks{E-mail address:
        eiltan@heraklit.physics.metu.edu.tr}
 \\
        Physics Department, Middle East Technical University \\
        Ankara, Turkey\\}

\date{}

\begin{document}
\setlength{\baselineskip}{24pt}
\maketitle
\setlength{\baselineskip}{7mm}
\begin{abstract}
We study the differential Branching ratio and $CP$ asymmetry for the 
exclusive decay  $B\rightarrow K^* l^+ l^-$ in the three Higgs doublet 
model with additional global $O(2)$ symmetry in the Higgs sector. We 
analyse dilepton mass square $q^2$ dependency of the these quantities. 
Further, we study the effect of new parameter of the global symmetry in 
the Higgs sector on the differential branching ratio and $CP$ asymmetry.
We see that there exist an enhancement in the branching ratio and  
a considerable $CP$ violation for the relevant process. In addition to this, 
we realize that fixing dilepton mass gives information about the sign of the
Wilson coefficient $C_7^{eff}$.Therefore, the future measurements of the 
CP asymmetry for $B\rightarrow K^* l^+ l^-$ decay will give a powerful 
information about the sign of Wilson coefficient $C_{7}^{eff}$ and new 
physics beyond the SM.
\end{abstract} 
\thispagestyle{empty}
\newpage
\setcounter{page}{1}

\section{Introduction}
Measurement of the physical quantities for rare B-decays provides sensitive
tests for the Standard model (SM) and it plays an important role in the
determination of the parameters, such as Cabbibo-Kobayashi-Maskawa (CKM) 
matrix elements, leptonic decay constants, etc., since these decays are 
induced by flavor changing neutral currents (FCNC) at loop level in the SM. 
Further, they give a comprehensive information in the search of  the physics 
beyond the SM, such as, two Higgs Doublet model (2HDM), Minimal Supersymmetric 
extension of the SM (MSSM) \cite{Hewett}, etc.
From the experimental point of view, the physical quantities, like Branching 
ratio ($Br$), $CP$ asymmetry ($A_{CP}$), forward backward asymmetry ($A_{FB}$), 
in rare B-decays have an outstanding role to obtain restrictions for the free 
parameters of the theory under consideration. 

Among the rare B-decays, $B\rightarrow K^*l^+ l^-$, induced by the inclusive 
process $b\rightarrow s l^+ l^-$, has a large $Br$ in the framework of the SM 
and it can be measured in future experiments. Therefore, the study on this 
process becomes attractive. In the literature, there are various studies on 
these decays in the framework of the SM, 2HDM and MSSM  
\cite{R4}- \cite{alil3}.

$CP$ violating effect is another physical quantity which can give information 
about the free parameters of the model. For $B\rightarrow K^* l^+ l^-$ decay, 
$CP$ violation almost vanishes in the framework of the SM, since the matrix 
element of the inclusive decay $b\rightarrow s l^+ l^-$, inducing 
$B\rightarrow K^* l^+ l^-$ process, contains only $V_{tb}V_{ts}^*$, due to the unitarity of CKM, 
$V_{ib}V_{is}^*=0\,\,, (i=u,c,t)$ and smallness of the term $V_{ub}V_{us}^*$.
This problem was studied in the general 2HDM, so called model III, which has 
a new source for $CP$ violation \cite{eril2}. In that work, the Yukawa 
couplings are taken complex and extra phase angles appear. These new 
parameters produce a considerable $CP$ violation effect for the decay under
consideration. 

In this work, we study $Br$ and $A_{CP}$ for the exclusive decay 
$B\rightarrow K^* l^+ l^-$ in the framework of the three Higgs doublet
model. Similar to the model III, complex Yukawa couplings are possible $CP$
violating sources. However, in the 3HDM, the number of free parameters
are large compared to that of 2HDM since the Higss sector is extended. We 
solve this problem by introducing a new global $O(2)$ symmetry in the Higgs 
sector. 

Even if the theoretical analysis of exclusive decays is more complicated 
due to the hadronic form factors, the experimental investigation of them  
is  easier compared those of inclusive ones. Therefore, this work is devoted
to the study of the exclusive $B\rightarrow K^* l^+ l^-$ decay. 

The paper is organized as follows:
In Section 2, we present our theoretical work based on 3HDM and the matrix 
element for the inclusive $b\rightarrow s l^+ l^- \,\, (l=e,\mu)$ 
decay in this model. In Section 3, we calculate $Br$ and $A_{CP}$ for the 
exclusive decay $B\rightarrow K^* l^+ l^-$. Section 4 is devoted to 
discussion and our conclusions. In Appendix, we give some theoretical results 
for the 3HDM and explicit forms of the necessary functions appear in the text.
\section{The inclusive $b\rightarrow s l^+ l^-$ decay in the 
framework of 3HDM    } 
In this section, we will derive the matrix element of the inclusive decay 
$b\rightarrow s l^+ l^-$ ($l=e,\mu$), which induces the exclusive 
$B\rightarrow K^* l^+ l^-$ process, in the framework of the 3HDM. 
We start with the general Yukawa interaction, 
\begin{eqnarray}
{\cal{L}}_{Y}&=&\eta^{U}_{ij} \bar{Q}_{i L} \tilde{\phi_{1}} U_{j R}+
\eta^{D}_{ij} \bar{Q}_{i L} \phi_{1} D_{j R}+
\xi^{U}_{ij} \bar{Q}_{i L} \tilde{\phi_{2}} U_{j R}+
\xi^{D}_{ij} \bar{Q}_{i L} \phi_{2} D_{j R} \nonumber \\
&+&
\rho^{U}_{ij} \bar{Q}_{i L} \tilde{\phi_{3}} U_{j R}+
\rho^{D}_{ij} \bar{Q}_{i L} \phi_{3} D_{j R}
 + h.c. \,\,\, ,
\label{lagrangian}
\end{eqnarray}
where $L$ and $R$ denote chiral projections $L(R)=1/2(1\mp \gamma_5)$,
$\phi_{i}$ for $i=1,2,3$, are three scalar doublets and  
$\eta^{U,D}_{ij}$, $\xi^{U,D}_{ij}$, $\rho^{U,D}_{ij}$ are
the Yukawa matrices having complex entries, in general. Now, we choose 
scalar Higgs doublets such that the first one describes only the SM part 
and last two carry the information about new physics beyond the SM: 
\begin{eqnarray}
\phi_{1}=\frac{1}{\sqrt{2}}\left[\left(\begin{array}{c c} 
0\\v+H^{0}\end{array}\right)\; + \left(\begin{array}{c c} 
\sqrt{2} \chi^{+}\\ i \chi^{0}\end{array}\right) \right]\, , 
\nonumber \\ \\
\phi_{2}=\frac{1}{\sqrt{2}}\left(\begin{array}{c c} 
\sqrt{2} H^{+}\\ H^1+i H^2 \end{array}\right) \,\, ,\,\, 
\phi_{3}=\frac{1}{\sqrt{2}}\left(\begin{array}{c c} 
\sqrt{2} F^{+}\\ H^3+i H^4 \end{array}\right) \,\, ,\nonumber
\label{choice}
\end{eqnarray}
with the vacuum expectation values,  
\begin{eqnarray}
<\phi_{1}>=\frac{1}{\sqrt{2}}\left(\begin{array}{c c} 
0\\v\end{array}\right) \,  \, ; 
<\phi_{2}>=0 \,\, ; <\phi_{3}>=0\,\, . 
\label{choice2}
\end{eqnarray}
Note that, the similar choice was done in the literature for the general 
2HDM \cite{soni}. The Yukawa interaction due to the new physics beyond the
SM  part is responsible for 
the Flavor Changing (FC) interactions and it can be written as 
\begin{eqnarray}
{\cal{L}}_{Y,FC}=
\xi^{U}_{ij} \bar{Q}_{i L} \tilde{\phi_{2}} U_{j R}+
\xi^{D}_{ij} \bar{Q}_{i L} \phi_{2} D_{j R}
+\rho^{U}_{ij} \bar{Q}_{i L} \tilde{\phi_{3}} U_{j R}+
\rho^{D}_{ij} \bar{Q}_{i L} \phi_{3} D_{j R} + h.c. \,\, .
\label{lagrangianFC}
\end{eqnarray}
Here, the couplings  $\xi^{U,D}$ and $\rho^{U,D}$ for the charged FC 
interactions are 
\begin{eqnarray}
\xi^{U}_{ch}&=& \xi_{N} \,\, V_{CKM} \nonumber \,\, ,\\
\xi^{D}_{ch}&=& V_{CKM} \,\, \xi_{N}  \nonumber \,\, , \\
\rho^{U}_{ch}&=& \rho_{N} \,\, V_{CKM} \nonumber \,\, ,\\
\rho^{D}_{ch}&=& V_{CKM} \,\, \rho_{N} \,\, ,
\label{ksi1} 
\end{eqnarray}
and
\begin{eqnarray}
\xi^{U,D}_{N}&=&(V_L^{U,D})^{-1} \xi^{U,D}\, V_R^{U,D}\,\, , \nonumber \\
\rho^{U,D}_{N}&=&(V_L^{U,D})^{-1} \rho^{U,D}\, V_R^{U,D}\,\, ,
\label{ksineut}
\end{eqnarray}
where the index "$N$" in $\xi^{U,D}_{N}$ denotes the word "neutral". 

At this stage, we obtain the effective Hamiltonian for the inclusive process 
$b\rightarrow s l^+ l^-$ by matching the full theory with the effective 
low energy one at the high scale $\mu$. In this calculation, there are 
additional charged Higgs effects coming from the new charged Higgs particles 
$F^{\pm}$ (see eqs. (\ref{CoeffH}) and (\ref{CoeffH2})). Fortunately, 
$F^{\pm} -quark-quark$ interaction  is the same as $H^{\pm}-quark-quark$ 
one  except new Yukawa couplings and they give additional contributions 
to the Wilson coefficients withouth changing the operator basis $O_i(\mu)$ 
(see \cite{alil3}). The Wilson coefficients are evaluated  from $\mu$ down 
to the lower scale $\mu\sim O(m_{b})$ using the renormalization group 
equations. Here the problem is to choose the high scale. In the literature, 
this scale is taken as the mass of charged Higgs, $\mu=m_{H^{\pm}}$, in the 
2HDM, since the evaluation from $\mu=m_{H^{\pm}}$ to $\mu=m_{W}$  gives 
negligible contribution to the Wilson coefficients( see \cite{alil1}). 
In our case, there is a new charged Higgs $F^{\pm}$ and its mass 
$m_{F^{\pm}}$ can be greater compared to $m_{H^{\pm}}$. However, by
introducing a new symmetry in the Higgs sector, we can take that masses of 
$F^{\pm}$ and $H^{\pm}$ are the same. Before starting with this
discussion, we would like to present the effective Hamiltonian, obtained by 
integrating out the heavy degrees of freedom, here, $t$ quark, $W^{\pm}$, 
$H^{\pm}$, $F^{\pm}$, $H^{1}$, $H^{2}$, $H^{3}$ and $H^{4}$  
where $H^{\pm}$, $F^{\pm}$ and $H^{1}$,$H^{2}$, $H^{3}$,$H^{4}$ denote 
charged and neutral Higgs bosons respectively: 
\begin{eqnarray}
{\cal{H}}_{eff}=-4 \frac{G_{F}}{\sqrt{2}} V_{tb} V^{*}_{ts} 
\sum_{i=1}^{12}(C_{i}(\mu) O_{i}(\mu)+C'_{i}(\mu) O'_{i}(\mu)) \, \, .
\label{hamilton}
\end{eqnarray}
In this equation, $O_{i}$ are current-current ($i=1,2,11,12$), penguin 
($i=1,...6$), magnetic penguin ($i=7,8$) and semileptonic ($i=9,10$) 
operators \cite{alil3, Grinstein2, misiak} and primed counterparts are their
flipped chirality partners \cite{alil3}. $C_{i}(\mu)$ and $C'_{i}(\mu)$ are 
Wilson coefficients renormalized at the scale $\mu$. 
The initial values of the Wilson coefficients in the SM model, 
$C_{i}^{(\prime)SM}(m_{W})$, can be found in  Appendix A. 
The additional contributions to the initial values of the Wilson 
coefficients, due to two new Higgs scalars are denoted by 
$C_{i}^{H}(m_{W})$ and for unprimed set of operators we have
\begin{eqnarray}
C^{H}_{1,\dots 6,11,12}(m_W)&=&0 \nonumber \, \, , \\
C_7^{H}(m_W)&=&\frac{1}{m_{t}^2} \,
(\bar{\xi}^{* U}_{N,tt}+\bar{\xi}^{* U}_{N,tc}
\frac{V_{cs}^{*}}{V_{ts}^{*}}) \, (\bar{\xi}^{U}_{N,tt}+\bar{\xi}^{U}_{N,tc}
\frac{V_{cb}}{V_{tb}}) F_{1}(y)\nonumber  \, \,  \\
&+&\frac{1}{m_t m_b} \, (\bar{\xi}^{* U}_{N,tt}+\bar{\xi}^{* U}_{N,tc}
\frac{V_{cs}^{*}}{V_{ts}^{*}}) \, (\bar{\xi}^{D}_{N,bb}+\bar{\xi}^{D}_{N,sb}
\frac{V_{ts}}{V_{tb}}) F_{2}(y)\nonumber  \, \,  \\ &+&
\frac{1}{m_{t}^2} \,(\bar{\rho}^{* U}_{N,tt}+\bar{\rho}^{* U}_{N,tc}
\frac{V_{cs}^{*}}{V_{ts}^{*}}) \, (\bar{\rho}^{U}_{N,tt}+\bar{\rho}^{U}_{N,tc}
\frac{V_{cb}}{V_{tb}}) F_{1}(y')\nonumber  \, \,  \\
&+&\frac{1}{m_t m_b} \, (\bar{\rho}^{* U}_{N,tt}+\bar{\rho}^{* U}_{N,tc}
\frac{V_{cs}^{*}}{V_{ts}^{*}}) \, (\bar{\rho}^{D}_{N,bb}+\bar{\rho}^{D}_{N,sb}
\frac{V_{ts}}{V_{tb}}) F_{2}(y')
\nonumber  \, \, , \\
C_8^{H}(m_W)&=&\frac{1}{m_{t}^2} \,
(\bar{\xi}^{* U}_{N,tt}+\bar{\xi}^{* U}_{N,tc}
\frac{V_{cs}^{*}}{V_{ts}^{*}}) \, (\bar{\xi}^{U}_{N,tt}+\bar{\xi}^{U}_{N,tc}
\frac{V_{cb}}{V_{tb}})G_{1}(y)
\nonumber  \, \,  \\
&+&\frac{1}{m_t m_b} \, (\bar{\xi}^{* U}_{N,tt}+\bar{\xi}^{* U}_{N,tc}
\frac{V_{cs}^{*}}{V_{ts}^{*}}) \, (\bar{\xi}^{D}_{N,bb}+\bar{\xi}^{U}_{N,sb}
\frac{V_{ts}}{V_{tb}}) G_{2}(y) \nonumber\, \,  \\
&+&
\frac{1}{m_{t}^2} \,
(\bar{\rho}^{* U}_{N,tt}+\bar{\rho}^{* U}_{N,tc}
\frac{V_{cs}^{*}}{V_{ts}^{*}}) \, (\bar{\rho}^{U}_{N,tt}+\bar{\rho}^{U}_{N,tc}
\frac{V_{cb}}{V_{tb}})G_{1}(y')
\nonumber  \, \,  \\
&+&\frac{1}{m_t m_b} \, (\bar{\rho}^{* U}_{N,tt}+\bar{\rho}^{* U}_{N,tc}
\frac{V_{cs}^{*}}{V_{ts}^{*}}) \, (\bar{\rho}^{D}_{N,bb}+\bar{\rho}^{U}_{N,sb}
\frac{V_{ts}}{V_{tb}}) G_{2}(y') \nonumber\, \, , \\
C_9^{H}(m_W)&=&\frac{1}{m_{t}^2} \,
(\bar{\xi}^{* U}_{N,tt}+\bar{\xi}^{* U}_{N,tc}
\frac{V_{cs}^{*}}{V_{ts}^{*}}) \, (\bar{\xi}^{U}_{N,tt}+\bar{\xi}^{U}_{N,tc}
\frac{V_{cb}}{V_{tb}}) H_{1}(y)
\nonumber  \, \,  \\ &+&
\frac{1}{m_{t}^2} \,
(\bar{\rho}^{* U}_{N,tt}+\bar{\rho}^{* U}_{N,tc}
\frac{V_{cs}^{*}}{V_{ts}^{*}}) \, (\bar{\rho}^{U}_{N,tt}+\bar{\rho}^{U}_{N,tc}
\frac{V_{cb}}{V_{tb}}) H_{1}(y')
\nonumber  \, \, , \\
C_{10}^{H}(m_W)&=&\frac{1}{m_{t}^2} \,
(\bar{\xi}^{* U}_{N,tt}+\bar{\xi}^{* U}_{N,tc}
\frac{V_{cs}^{*}}{V_{ts}^{*}}) \, (\bar{\xi}^{U}_{N,tt}+\bar{\xi}^{U}_{N,tc}
\frac{V_{cb}}{V_{tb}}) L_{1}(y) \, \, \nonumber \\ 
&+&\frac{1}{m_{t}^2} \,
(\bar{\rho}^{* U}_{N,tt}+\bar{\rho}^{* U}_{N,tc}
\frac{V_{cs}^{*}}{V_{ts}^{*}}) \, (\bar{\rho}^{U}_{N,tt}+\bar{\rho}^{U}_{N,tc}
\frac{V_{cb}}{V_{tb}}) L_{1}(y') \, \, , 
\label{CoeffH}
\end{eqnarray}
and for primed set of operators, 
\begin{eqnarray}
C^{\prime H}_{1,\dots 6,11,12}(m_W)&=&0 \nonumber \, \, , \\
C^{\prime H}_7(m_W)&=&\frac{1}{m_t^2} \,
(\bar{\xi}^{* D}_{N,bs}\frac{V_{tb}}{V_{ts}^{*}}+\bar{\xi}^{* D}_{N,ss})
\, (\bar{\xi}^{D}_{N,bb}+\bar{\xi}^{D}_{N,sb}
\frac{V_{ts}}{V_{tb}}) F_{1}(y)
\nonumber  \, \, \\
&+& \frac{1}{m_t m_b}\, (\bar{\xi}^{* D}_{N,bs}\frac{V_{tb}}{V_{ts}^{*}}
+\bar{\xi}^{* D}_{N,ss}) \, (\bar{\xi}^{U}_{N,tt}+\bar{\xi}^{U}_{N,tc}
\frac{V_{cb}}{V_{tb}}) F_{2}(y)
\nonumber  \, \, \\ &+&
\frac{1}{m_t^2} \,
(\bar{\rho}^{* D}_{N,bs}\frac{V_{tb}}{V_{ts}^{*}}+\bar{\rho}^{* D}_{N,ss})
\, (\bar{\rho}^{D}_{N,bb}+\bar{\rho}^{D}_{N,sb}
\frac{V_{ts}}{V_{tb}}) F_{1}(y')
\nonumber  \, \, \\
&+& \frac{1}{m_t m_b}\, (\bar{\rho}^{* D}_{N,bs}\frac{V_{tb}}{V_{ts}^{*}}
+\bar{\rho}^{* D}_{N,ss}) \, (\bar{\rho}^{U}_{N,tt}+\bar{\rho}^{U}_{N,tc}
\frac{V_{cb}}{V_{tb}}) F_{2}(y')\, , \nonumber \\
C^{\prime H}_8 (m_W)&=&\frac{1}{m_t^2} \,
(\bar{\xi}^{* D}_{N,bs}\frac{V_{tb}}{V_{ts}^{*}}+\bar{\xi}^{* D}_{N,ss})
\, (\bar{\xi}^{D}_{N,bb}+\bar{\xi}^{D}_{N,sb}
\frac{V_{ts}}{V_{tb}}) G_{1}(y)
\nonumber  \, \, \\
&+&\frac{1}{m_t m_b} \, (\bar{\xi}^{* D}_{N,bs}\frac{V_{tb}}{V_{ts}^{*}}
+\bar{\xi}^{* D}_{N,ss}) \, (\bar{\xi}^{U}_{N,tt}+\bar{\xi}^{U}_{N,tc}
\frac{V_{cb}}{V_{tb}}) G_{2}(y) 
\nonumber \,\, \\ 
&+&\frac{1}{m_t^2} \,
(\bar{\rho}^{* D}_{N,bs}\frac{V_{tb}}{V_{ts}^{*}}+\bar{\rho}^{* D}_{N,ss})
\, (\bar{\rho}^{D}_{N,bb}+\bar{\rho}^{D}_{N,sb}
\frac{V_{ts}}{V_{tb}}) G_{1}(y')
\nonumber  \, \, \\
&+&\frac{1}{m_t m_b} \, (\bar{\rho}^{* D}_{N,bs}\frac{V_{tb}}{V_{ts}^{*}}
+\bar{\rho}^{* D}_{N,ss}) \, (\bar{\rho}^{U}_{N,tt}+\bar{\rho}^{U}_{N,tc}
\frac{V_{cb}}{V_{tb}}) G_{2}(y')\, , \nonumber \\
C^{\prime H}_9(m_W)&=&\frac{1}{m_t^2} \,
(\bar{\xi}^{* D}_{N,bs}\frac{V_{tb}}{V_{ts}^{*}}+\bar{\xi}^{* D}_{N,ss})
\, (\bar{\xi}^{D}_{N,bb}+\bar{\xi}^{D}_{N,sb}
\frac{V_{ts}}{V_{tb}}) H_{1}(y)\nonumber  \, \,  \\ 
&+&\frac{1}{m_t^2} \,
(\bar{\rho}^{* D}_{N,bs}\frac{V_{tb}}{V_{ts}^{*}}+\bar{\rho}^{* D}_{N,ss})
\, (\bar{\rho}^{D}_{N,bb}+\bar{\rho}^{D}_{N,sb}
\frac{V_{ts}}{V_{tb}}) H_{1}(y')
\nonumber \,\, ,\\
C^{\prime H}_{10} (m_W)&=&\frac{1}{m_t^2} \,
(\bar{\xi}^{* D}_{N,bs}\frac{V_{tb}}{V_{ts}^{*}}+\bar{\xi}^{D}_{N,ss})
\, (\bar{\xi}^{D}_{N,bb}+\bar{\xi}^{D}_{N,sb}
\frac{V_{ts}}{V_{tb}}) L_{1}(y) \nonumber \\ &+&
\frac{1}{m_t^2} \,
(\bar{\rho}^{* D}_{N,bs}\frac{V_{tb}}{V_{ts}^{*}}+\bar{\rho}^{D}_{N,ss})
\, (\bar{\rho}^{D}_{N,bb}+\bar{\rho}^{D}_{N,sb}
\frac{V_{ts}}{V_{tb}}) L_{1}(y') \,\, ,
\label{CoeffH2}
\end{eqnarray}
where $y=m_t^2/m_{H^{\pm}}^2$ and $y'=m_t^2/m_{F^{\pm}}^2$.
In eqs.~(\ref{CoeffH}) and (\ref{CoeffH2}) we used the redefinition
\begin{eqnarray}
\xi^{U,D}=\sqrt{\frac{4 G_{F}}{\sqrt{2}}} \,\, \bar{\xi}^{U,D}\,\, .
\label{ksidefn}
\end{eqnarray}  
The explicit forms of the functions $F_{1(2)}(y)$, $G_{1(2)}(y)$, 
$H_{1}(y)$ and $L_{1}(y)$ can be found in Appedix A. 
In the calculations, we neglect the contributions of the neutral Higgs 
bosons to the Wilson coefficient $C_7^{eff}$ (see \cite {alil2}). Note 
that, the neutral Higgs bosons coming from $\phi_3$ give contribution to 
$C_7^{eff}$, including the Yukawa couplings  $\bar{\rho}^{D}_{N,bj}$ and 
$\bar{\rho}^{D}_{N,is}$ ($j=d,s$\, ;\, $i=d,s,b$), similar to the ones 
coming from $\phi_2$ \cite{alil2}.

In the 3HDM model, the Higgs sector is extended and this leads to an increase 
in the number of free parameters, namely, masses of new charged and neutral 
Higgs particles, new Yukawa couplings. In the Appendix B, we give the general 
gauge and CP invariant Higgs potential for the 3HDM and present the masses of 
charged and neutral Higgs particles. Now, our aim is to decrease the number 
of free parameters in the model under consideration. We consider three Higgs 
scalars as orthogonal vectors in a new space, which we call Higgs flavor 
space and we denote the Higgs flavor index by  "$m$", where $m=1,2,3$. 
We introduce a new global symmetry on the Higgs sector which keeps the 3HDM 
Lagrangian invariant. Let us take the following $O(2)$ transformation: 
\begin{eqnarray}
\phi'_{1}&=&\phi_{1}\nonumber \,\,,\\
\phi'_{2}&=&cos\,\alpha\,\, \phi_2+sin\,\alpha\,\, \phi_3 \,\, , \nonumber \\
\phi'_{3}&=&-sin\,\alpha\,\, \phi_2 + cos\,\alpha\,\, \phi_3\,\,,
\label{trans}
\end{eqnarray}
where $\alpha$ is the global parameter, which represents a rotation of 
the vectors $\phi_2$ and $\phi_3$ along the axis that $\phi_1$ lies, 
in the Higgs flavor space.  The kinetic term of the Lagrangian 
(see Appendix B) is invariant under this transformation. The invariance 
of the potential term can be obtained if the following conditions on the 
free parameters (eq. \ref{potential}) are satisfied:
\begin{eqnarray}
& & c_5=c_6\,\,\, , c_8=c_9\,\, , c_{11}=c_{12}\,\, , \nonumber \\
& & c_2=c_3=c_7=c_{10}=0 \,\, ,
\label{ci}
\end{eqnarray}
and we get 
\begin{eqnarray}
V(\phi_1, \phi_2,\phi_3 )&=&c_1 (\phi_1^+ \phi_1-v^2/2)^2+
c_4 [(\phi_1^+ \phi_1-v^2/2)^2+ \phi_2^+ \phi_2+\phi_3^+ \phi_3]^2
\nonumber \\ &+& 
c_5 [(\phi_1^+ \phi_1) (\phi_2^+ \phi_2 + \phi_3^+ \phi_3 )-
(\phi_1^+ \phi_2)(\phi_2^+ \phi_1)-(\phi_1^+ \phi_3)(\phi_3^+ \phi_1)]
\nonumber \\ &+& 
c_8( [Re(\phi_1^+ \phi_2)]^2 +[Re(\phi_1^+ \phi_3)]^2) +
c_{11} ([Im(\phi_1^+ \phi_2)]^2 +[Im(\phi_1^+ \phi_3)]^2) \nonumber \\
&+& c_{13} [Im(\phi_2^+ \phi_3)]^2 +c_{14}
\label{potential2}
\end{eqnarray}
Therefore, the masses of new particles are
\begin{eqnarray}
m_{F^\pm}=m_{H^\pm}=c_5 \frac{v^2}{2}\nonumber \,\, , \\
m_{H^3}=m_{H^1}=c_{8} \frac{v^2}{2}\nonumber \,\, , \\
m_{H^4}=m_{H^2}=c_{11} \frac{v^2}{2}\nonumber \,\, , \\
\label{mass2}
\end{eqnarray}
It is the first gain in decreasing the number of free parameters.
Now, we apply this transformation to the Yukawa Lagrangian 
(eq.(\ref{lagrangian})). This term is invariant if the transformed 
Yukawa matrices satisfy the expressions
\begin{eqnarray}
\bar{\xi}^{\prime U(D)}_{ij}&=& \bar{\xi}^{U (D)}_{ij} cos\, \alpha+
\bar{\rho}^{U(D)}_{ij} sin\, \alpha\,\, \nonumber ,\\
\bar{\rho}^{\prime U (D)}_{ij}&=&-\bar{\xi}^{U (D)}_{ij} sin\, \alpha+
\bar{\rho}^{U (D)}_{ij} cos\, \alpha \,\, .
\label{yuktr}
\end{eqnarray}
and therefore 
\begin{eqnarray}
(\bar{\xi}^{\prime U(D)})^+ \bar{\xi}^{\prime U (D) } +
(\bar{\rho}^{\prime U (D)})^+\bar{\rho}^{\prime U (D) }=
(\bar{\xi}^{U (D)})^+\bar{\xi}^{U (D)} +
(\bar{\rho}^{U (D)})^+ \bar{\rho}^{U (D) }\,\, , 
\label{yukinv}
\end{eqnarray}
which permits us to parametrize the Yukawa matrices $\bar{\xi}^{U(D)}$
and $\bar{\rho}^{U(D)}$ as 
\begin{eqnarray}
\bar{\xi}^{U (D)}=\epsilon^{U(D)} cos\,\theta \nonumber \,\, ,\\
\bar{\rho}^{U}=\epsilon^{U} sin\,\theta \nonumber \,\, ,\\ 
\bar{\rho}^{D}=i \epsilon^{D} sin\,\theta \,\, ,
\label{yukpar}
\end{eqnarray}
where $\epsilon^{U(D)}$ are real matrices satisfy the equation 
\begin{eqnarray}
(\bar{\xi}^{\prime U(D)})^+ \bar{\xi}^{\prime U (D) } +
(\bar{\rho}^{\prime U (D)})^+\bar{\rho}^{\prime U (D) }=
(\epsilon^{U(D)})^T \epsilon^{U(D)} 
\label{yukpareq}
\end{eqnarray}
Here $T$ denotes transpose operation. In eq. (\ref{yukpar}),  we take 
$\bar{\rho^{D}}$ complex to carry all $CP$ violating effects on the third 
Higgs scalar. 

Finally, we could reduce the number of free parameters, here the Yukawa 
matrices  $\bar{\xi}^{U,(D)}$ and $\bar{\rho}^{U,(D)}$, by connecting them 
by the expression given in eq. (\ref{yukpar}). Further, we take into account
only the Yukawa couplings $\bar{\xi}^{U}_{N, tt}$, $\bar{\xi}^{D}_{N,bb}$, 
$\bar{\rho}^{U}_{N,tt}$ and $\bar{\rho}^{D}_{N,bb}$, since we assume that 
the others are small due to the discussion given in \cite{alil3}.
Now, we rewrite the contributions of the charged Higgs particles
to the initial values of the Wilson coefficients as:
\begin{eqnarray}
C^{H}_{1,\dots 6,11,12}(m_W)&=&0 \nonumber \, \, , \\
C_7^{H}(m_W)&=&\frac{1}{m_{t}^2} \,
(\bar{\xi}^{* U}_{N,tt} \, \bar{\xi}^{U}_{N,tt}+
\bar{\rho *}^{U}_{N,tt}\, \bar{\rho}^{U}_{N,tt})F_{1}(y)
\nonumber  \\
&+&\frac{1}{m_t m_b} \, (\bar{\xi}^{* U}_{N,tt}\,\bar{\xi}^{D}_{N,bb}+
\bar{\rho}^{* U}_{N,tt}\,\bar{\rho}^{D}_{N,bb})F_{2}(y)
\nonumber  \, \, , \\ 
C_8^{H}(m_W)&=&\frac{1}{m_{t}^2} \,
(\bar{\xi}^{* U}_{N,tt} \,\bar{\xi}^{U}_{N,tt}+
\bar{\rho}^{* U}_{N,tt} \,\bar{\rho}^{U}_{N,tt})G_{1}(y)
\nonumber  \\
&+&\frac{1}{m_t m_b} \, (\bar{\xi}^{* U}_{N,tt} \bar{\xi}^{D}_{N,bb}+
\bar{\rho}^{* U}_{N,tt} \bar{\rho}^{D}_{N,bb})G_{2}(y) 
\nonumber\, \, , \\
C_9^{H}(m_W)&=&\frac{1}{m_{t}^2} \,
(\bar{\xi}^{* U}_{N,tt} \bar{\xi}^{U}_{N,tt}+
\bar{\rho}^{* U}_{N,tt} \bar{\rho}^{U}_{N,tt})H_{1}(y)
\nonumber  \, \, , \\ 
C_{10}^{H}(m_W)&=&\frac{1}{m_{t}^2} \,
(\bar{\xi}^{* U}_{N,tt} \bar{\xi}^{U}_{N,tt}+
\bar{\rho}^{* U}_{N,tt} \bar{\rho}^{U}_{N,tt})L_{1}(y)\,\, ,
\label{CoeffH3}
\end{eqnarray}
and neglect the primed coefficients since they include small Yukawa
couplings. Here, $\bar{\xi}^{U(D)}_{N, tt (bb)}$ and 
$\bar{\rho}^{U(D)}_{N, tt(bb)}$ can be obtained by using eq. (\ref{yukpar}).
Note that, with the replacements 
$\bar{\xi}^{* U}_{N,tt} \bar{\xi}^{U}_{N,tt}+
\bar{\rho}^{* U}_{N,tt} \bar{\rho}^{U}_{N,tt} \rightarrow 
\bar{\xi}^{* U}_{N,tt} \bar{\xi}^{U}_{N,tt}$
and $\bar{\xi}^{* U}_{N,tt} \bar{\xi}^{D}_{N,bb}+
\bar{\rho}^{* U}_{N,tt} \bar{\rho}^{D}_{N,bb}\rightarrow 
\bar{\xi}^{* U}_{N,tt} \bar{\xi}^{D}_{N,bb}$\,
we get the results for the general 2HDM (model III) \cite{alil3}.
By neglecting the primed ones , the initial values of the Wilson 
coefficients can be written as  
\begin{eqnarray}
C_i^{3HDM}(m_W)&=&C_i^{SM}(m_W)+C_i^{H}(m_W)\,\, ,
\label{Coef3HDM}
\end{eqnarray}
and using these initial values, we can calculate the coefficients 
$C_{i}^{3HDM}(\mu)$ at any lower scale with five quark effective theory, 
namely $u,c,d,s,b$. In the process under consideration the Wilson 
coefficients $C_7^{eff}(\mu)$, $C_9^{eff}(\mu)$ and $C_{10}(\mu)$ play the 
important role in the physical quantities and the others enter into 
expressions with operator mixing. Besides the perturbative part, there exist 
the long distance (LD) effects due to the real $\bar{c}c$ in the intermediate 
states, 
i.e. the cascade process $B\rightarrow K^* \psi_i \rightarrow K^* l^+ l^-$
where $i=1,..,6$. These effects appear in the Wilson coefficient $C_9^{eff}$
and using a Breit-Wigner form of the resonance propagator 
\cite{buras,zakharov}, they are added to the perturbative one coming from 
the $c\bar{c}$ loop: 
\begin{eqnarray}
C_9^{eff}(\mu)=C_9^{pert}(\mu)+ Y_{reson}\,\, ,
\label{C9efftot}
\end{eqnarray}
where $Y_{reson}$ in NDR scheme is defined as
\begin{eqnarray}
Y_{reson}&=&-\frac{3}{\alpha^2_{em}}\kappa \sum_{V_i=\psi_i}
\frac{\pi \Gamma(V_i\rightarrow ll)m_{V_i}}{q^2-m^2_{V_i}+i m_{V_i}
\Gamma_{V_i}} \nonumber \\
& & \left( 3 C_1(\mu) + C_2(\mu) + 3 C_3(\mu) + 
C_4(\mu) + 3 C_5(\mu) + C_6(\mu) \right).
\label{Yres}
\end{eqnarray}
In eqs. (\ref{Yres}), the phenomenological parameter $\kappa$ is taken as 
$\kappa=2.3$ \cite{ali}. The explicit forms of the perturbative parts of the 
Wilson coefficients $C_i(\mu)\,\,, i=1,...,9$  including NLO QCD corrections 
can be found in the literature \cite{alil3, ciuchini2,greub2}.

Finally, neglecting the strange quark mass and primed coefficients, 
the matrix element for $b \rightarrow s \ell^+\ell^-$ decay is obtained as:
\begin{eqnarray}
{\cal M}&=& - \frac{G_F \alpha_{em}}{2\sqrt 2 \pi} V_{tb} V^*_{ts} 
\Bigg\{ \, C_9^{eff}(\mu)\,
\bar s \gamma_\mu (1- \gamma_5) b \,\, \bar \ell \gamma^\mu \ell 
+ C_{10}(\mu) \, \bar s \gamma_\mu (1- \gamma_5) b\,\, 
\bar \ell \gamma^\mu \gamma_5 \ell \nonumber  \\
&-& 2 C^{eff}_7(\mu)\, \frac{m_b}{q^2}\, 
\bar s i \sigma_{\mu \nu}q^\nu (1+\gamma_5)  b
\,\, \bar \ell \gamma^\mu \ell \Bigg\}~\nonumber .
\label{matr}
\end{eqnarray}
\section{The exclusive $B\rightarrow K^* l^+ l^-$ decay} 
In this section, we study the Branching ratio (Br) and the CP asymmetry
($A_{CP}$) of the exclusive decay  $B\rightarrow K^* l^+ l^-$ in the 3HDM. 
Using the results for the matrix elements 
$ \la K^* \vel \bar s \gamma_\mu (1\pm \gamma_5) b \ver B \ra$, and
$\la K^* \vel \bar s i \sigma_{\mu \nu} q^\nu (1\pm\gamma_5) b \ver B \ra$
\cite{colangelo}, the hadronic matrix element of the 
$B\rightarrow K^* l^+ l^-$ decay is obtained as \cite{alsav2}:
\begin{eqnarray}
{\cal M} &=& -\frac{G \alpha_{em}}{2 \sqrt 2 \pi} V_{tb} V_{ts}^*  
\Bigg\{ \bar \ell \gamma^\mu
\ell \left[ 2 A \epsilon_{\mu \nu \rho \sigma} \epsilon^{* \nu} 
p_{K^*}^\rho q^\sigma + i
B_{1} \epsilon^*_\mu - i B_{2} ( \epsilon^* q) 
(p_{B}+p_{K^*})_\mu - 
i B_{3} (\epsilon^* q)q_\mu \right] \nonumber \\
&+& \bar \ell \gamma^\mu \gamma_5 \ell \left[ 2 C \epsilon_{\mu \nu \rho
\sigma}\epsilon^{* \nu} p_{K^*}^\rho q^\sigma + i D_{1} \epsilon^*_\mu - 
i D_{2} (\epsilon^* q) (p_{B}+p_{K^*})_\mu - i D_{3} (\epsilon^* q) 
q_\mu \right] \Bigg\}~,
\label{matr2}
\end{eqnarray}
where $\epsilon^{* \mu}$ is the polarization vector of $K^*$ meson, $p_{B}$ 
and $p_{K^*}$ are four momentum vectors of $B$ and $K^*$ mesons, 
$q=p_B-p_{K^*}$ and $A$, $C$, $B_{i}$, and  $D_{i}$ 
$i=1,2,3$  are the form factors of the relevant process. 
Their explicit forms can be found in the Appendix C.

Using eq.(\ref{matr2}), we get the double differential decay rate:  
\begin{eqnarray}
\frac{d \Gamma}{d q^2 dz} &=& \frac{G^2 \alpha_{em}^2
\vel V_{tb} V_{ts}^* \ver^2\lambda^{1/2} }{2^{12}
\pi^5 m_B} \Bigg\{ 2 \lambda m_B^4 \Bigg[ 
m_B^2 s ( 1+ z^2) \ga \vel A \ver ^2 +\vel C \ver ^2 \dr
\Bigg] \nonumber \\
&+& \frac{\lambda m_B^4}{2 r} \Bigg[ \lambda m_B^2 (1-z^2) \ga \vel 
B_{2} \ver ^2 + \vel D_{2} \ver ^2 \dr \Bigg]  \nonumber \\
&+& \frac{1}{2 r} \Bigg[ m_B^2 \left\{ \lambda (1- z^2) + 8 r s\right\}
\ga \vel B_{1} \ver^2 + \vel D_{1} \ver^2 \dr  
\nonumber \\
&-& 2 \lambda m_B^4
(1-r-s)(1-z^2)  \left\{ Re\ga  B_{1} B_{2}^*  \dr + 
Re \ga D_{1} D_{2}^*  \dr \right\} 
\Bigg]  \nonumber \\
&-& 8 m_B^4 s\lambda^{1/2} z \Bigg[ 
\left\{ Re\ga  B_{1} C^* \dr + Re\ga A D_{1}^*\dr 
\right\} \Bigg] \Bigg\}~,
\label{dddr}
\end{eqnarray}
where $z=cos \beta$\,, $\beta$ is the angle between the momentum of $\ell$ 
lepton and that of $B$ meson in the center of mass frame of the lepton 
pair, $\lambda = 1+r^2+s^2 -2 r - 2 s - 2 r s$, $r =
\frac{\ds{m_{K^*}^2}}{\ds{m_B^2}}$ and 
$s=\frac{\ds{q^2}}{\ds{m_B^2}}$.

$A_{CP}$ is another important physical quantity which
almost does not exist for the given process in the framework of the SM.
However, with the choice of the complex Yukawa couplings, it is possible 
that such asymmetry exists, in extended models like 2HDM \cite{eril2}. 
In our case, the model under consideration is the 3HDM with global $O(2)$ 
symmetry in the Higgs sector and  the possible source of CP violation 
comes from complex Yukawa couplings in the third Higgs doublet.
Using the definition of $A_{CP}$
\begin{eqnarray}
A_{CP}= \frac{\frac{d \Gamma (\bar{B_s}\rightarrow K^* e^+ e^-)}{dq^2 }-
\frac{d \Gamma (B_s\rightarrow \bar{K^*} e^+ e^-)}{dq^2 }}
{\frac{d \Gamma (\bar{B_s}\rightarrow K^* e^+ e^-)}{dq^2 }+
\frac{d \Gamma (B_s\rightarrow \bar{K^*} e^+ e^-)}{dq^2 }}\,\, .
\label{cpvio}
\end{eqnarray}
we get 
\begin{eqnarray}
A_{CP}=-2 Im (\lambda_2) \frac{Im (C_9^{eff}(\mu)) \,\, P_1(\mu)\, \, \Delta}
{Re(\lambda_2)[-2 (P_1(\mu)+2 P_2(\mu))\, Re (C_9^{eff}(\mu))\,
\Delta +\Omega}.
\label{Acp}
\end{eqnarray}
In eq. (\ref{Acp}) we use the same parametrization as in \cite{eril2}
\begin{eqnarray}
C_7^{eff}(\mu)=P_1(\mu) \, \lambda_2 + P_2(\mu)\,\,,
\label{param}
\end{eqnarray}
where $\lambda_2$ is
\begin{eqnarray}
\lambda_2=\frac{1}{m_t\, m_b} \bar{\epsilon}^{U}_{N,tt} 
\bar{\epsilon}^{D}_{N, bb}
(cos^2\, \theta+ i sin^2 \, \theta)
\label{lambda2}
\end{eqnarray}
Here the functions $P_1(\mu)$ and $P_2(\mu)$ can be written as the 
combinations of LO and NLO part, namely,
\begin{eqnarray}
P_1(\mu)&=&P_1^{LO}(\mu)+P_1^{NLO}(\mu)\nonumber \,\, ,\\
P_2(\mu)&=&P_2^{LO}(\mu)+P_2^{NLO}(\mu)\,\, , 
\label{P12}
\end{eqnarray}
and 
\begin{eqnarray}
P_1^{LO}(\mu)&=&\eta^{16/23}
F_2(y)+\frac{8}{3}(\eta^{14/23}-\eta^{16/23})\,G_2(y)\nonumber \\
P_2^{LO}(\mu)&=&\eta^{16/23}
[C_7^{SM}(m_W)+\frac{|\bar{\epsilon}^{U}_{N,tt}|^2}{m_t^2}  F_1(y)]\nonumber \\
&+&\frac{8}{3}(\eta^{14/23}-\eta^{16/23})
[C_8^{SM}(m_W)+\frac{|\bar{\epsilon}^{U}_{N,tt}|^2}{m_t^2}  G_1(y)]\nonumber \\
&+&Q_d (C_5^{LO}(\mu)+N_c \, C_6^{LO}(\mu))+
Q_u (\frac{m_c}{m_b} C_{12}^{LO}(\mu)+N_c \frac{m_c}{m_b} C_{11}^{LO}(\mu))
\nonumber\\
&+& C_2(m_W) \sum_{i=1}^{8} h_i \eta^{a_i}\,\, ,
\label{P12LO}
\end{eqnarray}
where $\eta=\frac{\alpha_s(\mu)}{\alpha_s(m_W)}$, $h_i$ and $a_i$ are
numbers appear during the evaluation \cite{buras}.
$P_1^{NLO}(\mu)$ is the coefficient of $\lambda_2$ in the expression 
$\frac{\alpha_s(\mu)}{4 \pi} C_7^{(1)\, 3HDM}(\mu)$ and $P_2^{NLO}(\mu)$ 
is obtained by setting $\lambda_2=0$ in the 
same expression. The functions $\Delta$ and $\Omega$ are defined 
as
\begin{eqnarray}
\Delta&=& -\frac{T_2 s}{3 q^2 r (1+\sqrt r)}\Bigg \{ A_2 \lambda (-1-3r+s)+
A_1 (1+\sqrt r)^2(\lambda-12 r (r-1))\Bigg \} \nonumber \\ 
&+& \frac{T_3 \lambda}{3 m_B^2 r (1+\sqrt r)(r-1)}\Bigg 
\{ A_2 \lambda + A_1 (1+\sqrt r)^2 (-1+r+s))\Bigg \}
-\frac{8 T_1 V s}{3 q^2 (1+\sqrt r)}\lambda  \nonumber \,\, ,\\ \\
\Omega&=& \frac{|C_9^{eff}|^2+|C_{10}|^2}{6 m_b m_B (1+\sqrt r)^2 r}
\Bigg \{ 2 A_1 A_2 \lambda (1+\sqrt r)^2 (-1+r+s)+
A_1^2 (1+\sqrt r)^4 (\lambda+12 r s)\nonumber \\ 
&+& \lambda^2 A_2^2+8 \lambda r s V^2 \Bigg \} \nonumber \\ 
&+& 8 (P_1(\mu)+P_2(\mu)) P_2(\mu) \Bigg \{ \frac{8 \lambda m_b m_B}{3 q^4} 
T_1^2 s + \frac{m_b m_B}{3 q^4 r} T_2^2 [\lambda (-4 r+s)+ 12 r (r-1)^2]\, s
\nonumber \\ 
&+& \frac{m_b }{3 m_B^3 r (-1+r)^2} \lambda^2 T_3^2 \nonumber \\
&+& \frac{2 \lambda m_b }{3 m_B q^2r (-1+r)} s (1-s+3 r)T_2 T_3 \Bigg \}
\label{DelOme}
\end{eqnarray}
Here the form factors $A_1$, $A_2$, $T_{1}$, $T_2$, $T_3$ and $V$
can be found in the Appendix C.
\section{Discussion}
In the general 3HDM model, there are many free parameters, such as 
masses of charged and neutral Higgs bosons, complex Yukawa couplings, 
$\xi_{ij}^{U,D}$,$\rho_{ij}^{U,D}$ where $i,j$ are quark flavor indices. 
The additional global $O(2)$ symmetry in the Higgs flavor space   
forces that the masses of new charged Higgs particles to be the same.
Further, the masses of the new neutral Higgs particles in the second doublet 
$\phi_2$, are the same as those of the corresponding ones in the   
third doublet  $\phi_3$. This symmetry also connects the Yukawa matrices  
in the second and third doublet (see eq. (\ref{yukpar})). The Yukawa 
couplings, which are entries of Yukawa matrices, can be restricted using 
the experimental measurements. In our calculations, we neglect all Yukawa 
couplings except $\bar{\xi}^{U}_{N,tt}$ and $\bar{\xi}^{D}_{N,bb}$ 
$\bar{\rho}^{U}_{N,tt}$ and $\bar{\rho}^{D}_{N,bb}$
by respecting the CLEO measurement announced recently  
\cite{cleo2}, 
\begin{eqnarray}
Br (B\rightarrow X_s\gamma)= (3.15\pm 0.35\pm 0.32)\, 10^{-4} \,\, ,
\label{br2}
\end{eqnarray}

This section is devoted to the study of the $q^2$ dependencies of $Br$ and 
$A_{CP}$ for the decay $B\rightarrow K^* l^+ l^-$, for the selected 
parameters of the 3HDM ($\bar{\epsilon}^{U}_{N,tt}$,  
$\bar{\epsilon}^{D}_{N,bb}$ and the angle $\theta$) with $O(2)$ symmetry 
in the Higgs sector. In our analysis, we restricted $|C_7^{eff}|$ in the 
region $0.257 \leq |C_7^{eff}| \leq 0.439$, coming from CLEO measurement 
\cite{cleo2}, where upper and lower limits were calculated in \cite{alil1}
following the procedure given in \cite{gudalil}. This restriction allows us 
to define a constraint region for the parameter $\bar{\epsilon}^{U}_{N,tt}$ 
in terms of $\bar{\epsilon}^{D}_{N,bb}$ and $\theta$.
Our numerical calculations based on this restriction and the constraint for
the angle $\theta$ due to the experimental upper limit of neutron electric
dipole moment, namely  $d_n<10^{-25}\hbox{e$\cdot$cm}$ which places
a upper bound on the couplings with the expression: 
$\frac{1}{m_t m_b} (\bar{\epsilon}^{U}_{N,tt}\,
\bar{\epsilon}^{* D}_{N,bb})sin^2\,\theta < 1.0$ 
for $M_{H^\pm}\approx 200$ GeV \cite{david}. Throughout these calculations, 
we take the charged Higgs mass $m_{F^{\pm}}=m_{H^{\pm}}=400\, GeV$, the 
scale $\mu=m_b$ and we use the input values given in Table (\ref{input}).  
\begin{table}[h]
        \begin{center}
        \begin{tabular}{|l|l|}
        \hline
        \multicolumn{1}{|c|}{Parameter} & 
                \multicolumn{1}{|c|}{Value}     \\
        \hline \hline
        $m_c$                   & $1.4$ (GeV) \\
        $m_b$                   & $4.8$ (GeV) \\
        $\alpha_{em}^{-1}$      & 129           \\
        $\lambda_t$            & 0.04 \\
        $m_{B_d}$             & $5.28$ (GeV) \\
        $\Gamma_{tot} (B_d)$     & $3.96\, 10^{-3}$\\
        $m_{t}$             & $175$ (GeV) \\
        $m_{W}$             & $80.26$ (GeV) \\
        $m_{Z}$             & $91.19$ (GeV) \\
        $\Lambda_{QCD}$             & $0.214$ (GeV) \\
        $\alpha_{s}(m_Z)$             & $0.117$  \\
        $sin\theta_W$             & $0.2325$  \\
        \hline
        \end{tabular}
        \end{center}
\caption{The values of the input parameters used in the numerical
          calculations.}
\label{input}
\end{table}

\begin{figure}[htb]
\vskip -3.0truein
\centering
\epsfxsize=6.8in
\leavevmode\epsffile{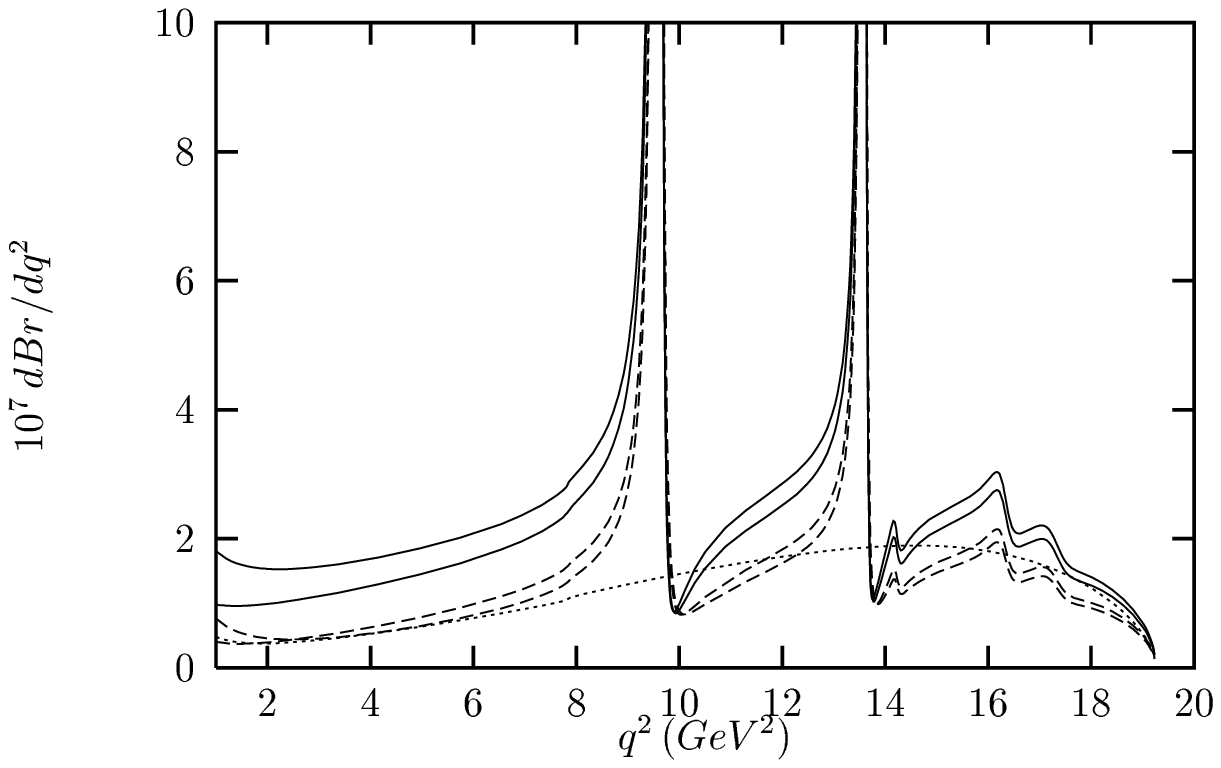}
\vskip -3.0truein
\caption[]{Differential $Br$ as a function of  $q^2$ for
$sin\,\theta =0.5$, $\bar{\epsilon}_{N,bb}^{D}=40\, m_b$ in the region 
$|r_{tb}|<1$, at the scale $\mu=m_b$, including LD effects. Here 
differential $Br$ is restricted in the region bounded by solid lines for 
$C_7^{eff}>0$ and by  dashed lines for $C_7^{eff}<0$. Dotted line represents 
the SM result withouth LD effects.}  
\label{brsintet05}
\end{figure}

\begin{figure}[htb]
\vskip -3.0truein
\centering
\epsfxsize=6.8in
\leavevmode\epsffile{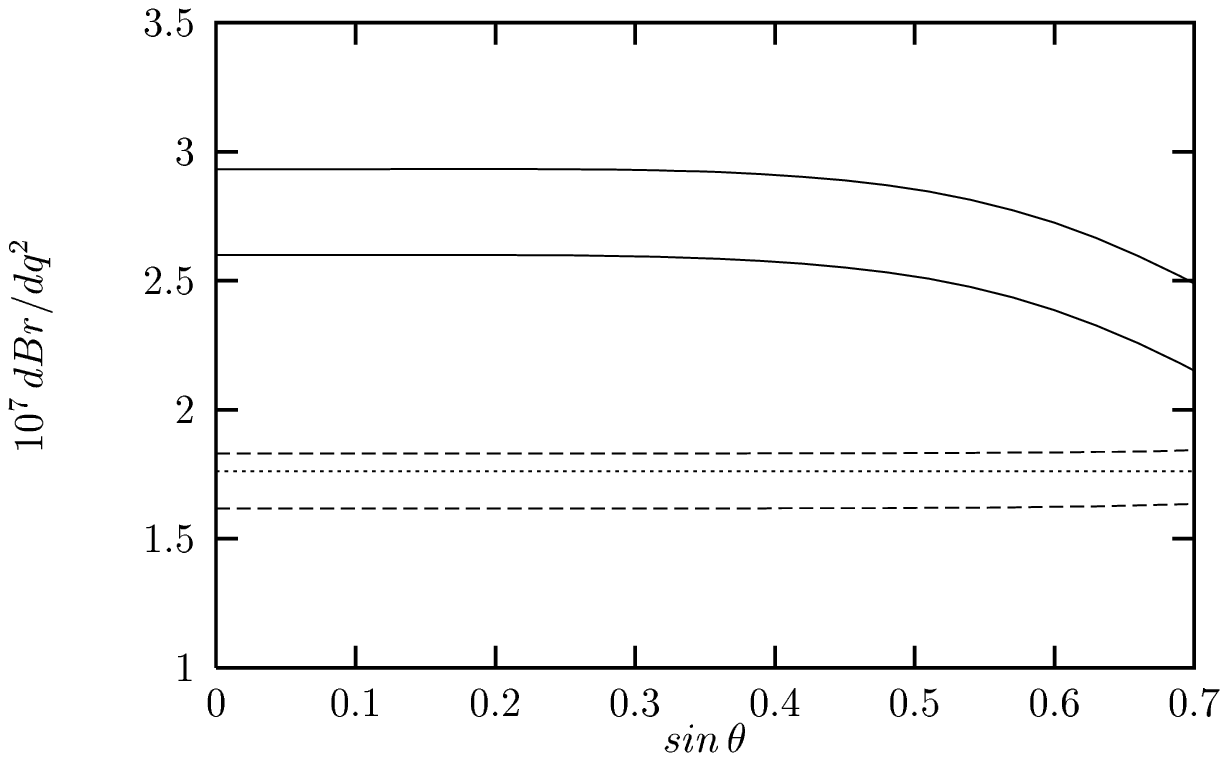}
\vskip -3.0truein
\caption[]{Differential $Br$ as a function of  $sin\,\theta$ for  
$q^2=12\, GeV^2$, $\bar{\epsilon}_{N,bb}^{D}=40\, m_b$ in the region 
$|r_{tb}|<1$, at the scale $\mu=m_b$, including LD effects. Here 
differential $Br$ is restricted in the region bounded by 
solid lines for $C_7^{eff}>0$ and by  dashed lines for $C_7^{eff}<0$.
Dotted line represents the SM result withouth LD effects.}
\label{brq212}
\end{figure}

In  fig.~\ref{brsintet05}, we plot the differential $Br$ of the decay 
$B\rightarrow K^* l^+ l^-$ with respect to the dilepton mass $q^2$ for 
$\bar{\epsilon}_{N,bb}^{D}=40\, m_b$, $sin\,\theta=0.5$ and charged Higgs 
mass $m_{H^{\pm}}=400\, GeV$ at the scale $\mu=m_b$. This figure represents 
the case where the ratio 
$|r_{tb}|=|\frac{\bar{\epsilon}_{N,tt}^{U}}{\bar{\epsilon}_{N,bb}^{D}}| <1.$
Here the differential $Br$ lies in the region bounded by solid lines for 
$C_7^{eff} > 0$ and by dashed  lines for  $C_7^{eff} < 0$. It is shown that 
there is an  enhancement for $C_7^{eff} > 0$ case compared to the SM (dotted
line). Further, the restriction region of the differential $Br$ for
$C_7^{eff} > 0$ case is broader than the one for $C_7^{eff} < 0$. 
Fig. \ref{brq212} is devoted the dependence of the differential $Br$ to 
$sin\,\theta$ for $|r_{tb}|<1$, $\bar{\epsilon}_{N,bb}^{D}=40\, m_b$, 
$q^2=12\, GeV^2$ and charged Higgs mass $m_{H^{\pm}}=400\, GeV$ at the 
scale $\mu=m_b$. Here, the differential $Br$ lies in the region between solid 
lines for $C_7^{eff} > 0$ and   lies in the region between dashed lines for
$C_7^{eff} < 0$. For $C_7^{eff} > 0$, there is a weak dependence to 
$sin\,\theta$ especially for $sin\,\theta < 0.5$. For $C_7^{eff} < 0$ this 
dependence almost vanishes. Now, we present the $Br$ for three different 
phase angles ($sin\theta=0.1,\,0,5,\,0.9$) in two different dilepton mass 
regions (Table \ref{Table2}),
\begin{table}[h]
\small{    \begin{center}
    \begin{tabular}{|c|c|c|c|c|}
    \hline
    \hline \hline
    $sin\theta$   &$C_7^{eff}>0$&  $C_7^{eff}<0$ &$q^2$ regions\\
\hline \hline
    $0.1$            
&$1.80\, 10^{-6}\leq  Br \leq 2.21\, 10^{-6} $ 
&$0.95\, 10^{-6}\leq Br\leq 1.07\, 10^{-6} $& I  \\
    \hline
&$0.96\, 10^{-6}\leq Br\leq 1.07\, 10^{-6} $ 
&$0.66\, 10^{-6}\leq Br\leq 0.72\, 10^{-6} $ & II \\
    \hline \hline

    $0.5$            
&$1.72\, 10^{-6}\leq Br \leq 2.12\, 10^{-6} $ 
&$0.94\, 10^{-6}\leq Br \leq 1.06\, 10^{-6} $& I  \\
    \hline
&$0.94\, 10^{-6}\leq Br \leq 1.05\, 10^{-6} $ 
&$0.66\, 10^{-6}\leq Br \leq 0.72\, 10^{-6} $ & II \\
    \hline \hline
    
    $0.9$             
&$1.04\, 10^{-6}\leq  Br \leq 1.26\, 10^{-6} $ 
&$1.04\, 10^{-6}\leq  Br \leq 1.05\, 10^{-6} $& I  \\
    \hline 
&$0.71\, 10^{-6}\leq  Br \leq 0.77\, 10^{-6} $ 
&$0.69\, 10^{-6}\leq  Br \leq 0.71\, 10^{-6} $ & II \\
    \hline \hline
        \end{tabular}
        \end{center} }
\caption{ $Br$ for regions I 
( $1\, GeV\leq \sqrt q^2 \leq m_{J/\psi}-20\, MeV$ ) and II 
($m_{J/\psi}+20\, MeV \leq \sqrt q^2 \leq m_{\psi'}-20\, MeV$ ) }
\label{Table2}
\end{table}

\begin{figure}[htb]
\vskip -3.0truein
\centering
\epsfxsize=6.8in
\leavevmode\epsffile{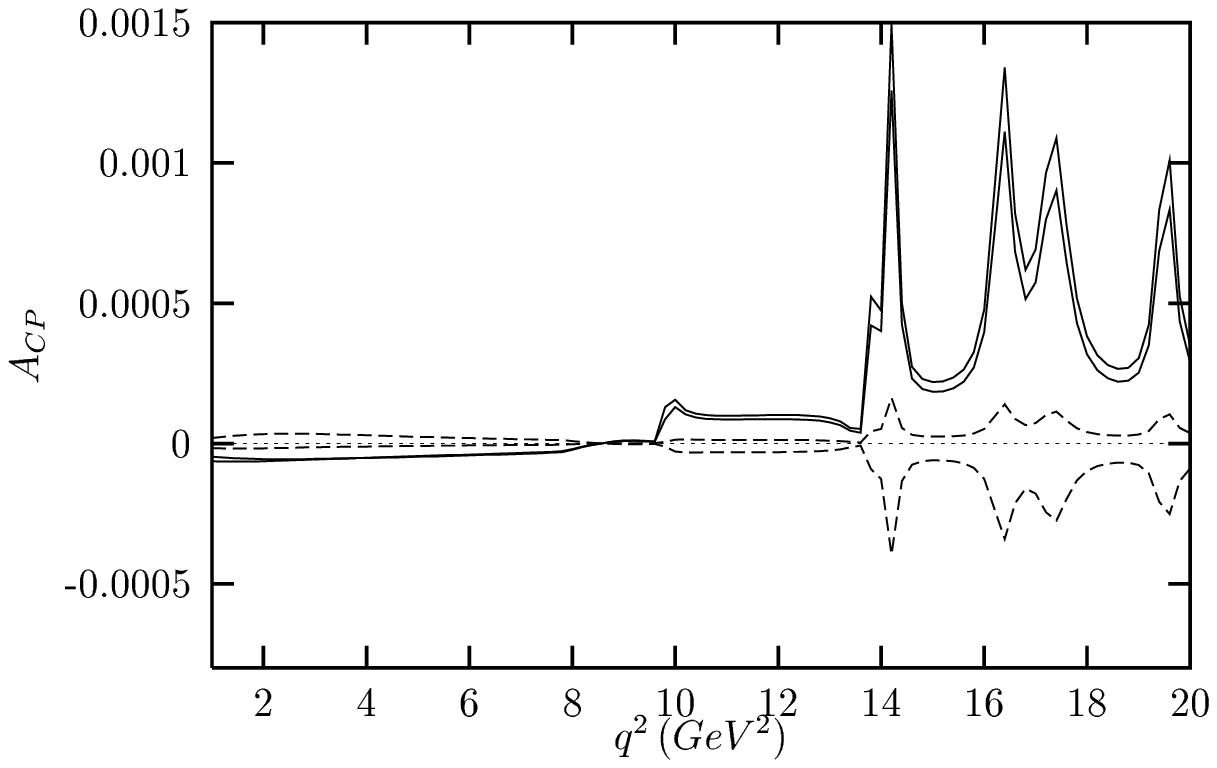}
\vskip -3.0truein
\caption[]{$A_{CP}$ as a function of  $q^2$ for $sin\,\theta =0.1$, 
$\bar{\epsilon}_{N,bb}^{D}=40\, m_b$ in the region 
$|r_{tb}|<1$, at the scale $\mu=m_b$, including LD effects. Here 
$A_{CP}$ is restricted in the region bounded by solid lines for 
$C_7^{eff}>0$ and by  dashed lines for $C_7^{eff}<0$ .}
\label{cpsintet01}
\end{figure}

\begin{figure}[htb]
\vskip -3.0truein
\centering
\epsfxsize=6.8in
\leavevmode\epsffile{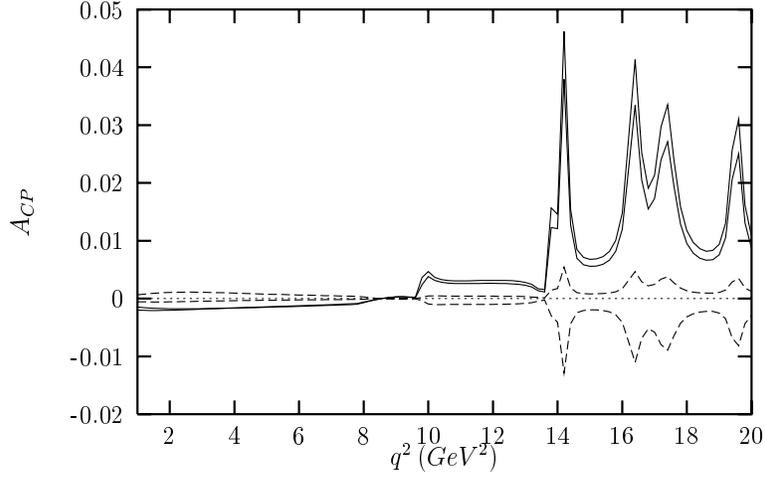}
\vskip -3.0truein
\caption[]{The same as Fig \ref{cpsintet01}, but for $sin\,\theta =0.5$.}
\label{cpsintet05}
\end{figure}

In  figs.~\ref{cpsintet01} (\ref{cpsintet05}) we plot $A_{CP}$ of the 
decay $B\rightarrow K^* l^+ l^-$ with respect to the dilepton mass square, 
$q^2$, for $\bar{\epsilon}_{N,bb}^{D}=40\, m_b$, 
$sin\,\theta=0.1$ ($sin\,\theta=0.5$) in the case where the ratio 
$|r_{tb}| < 1.$ For $C_7^{eff} > 0$, $A_{CP}$ is restricted in the region 
bounded by solid lines and for $C_7^{eff} < 0$ it lies between dashed
lines. $A_{CP}$ changes sign almost at the $q^2$ value $q^2 \sim 9\, GeV^2$ 
for $C_7^{eff} > 0$ case. However, for $C_7^{eff} < 0$, it can have both 
signs for any $q^2$ value. $A_{CP}$ enhances strongly with increasing 
value of $sin\,\theta$. For completeness, we present the average value 
of $A_{CP}$ for three different  phase angles ($sin\theta=0.1,\,0,5,\,0.9$) 
in two different dilepton mass regions (Table \ref{Table3}),
\begin{table}[h]
\small{    \begin{center}
    \begin{tabular}{|c|c|c|c|c|}
    \hline
    \hline \hline
    $sin\theta$   &$C_7^{eff}>0$&  $C_7^{eff}<0$ &$q^2$ regions\\
\hline \hline
    $0.1$            
&$-3.35\, 10^{-4}\leq \bar{A}_{CP}\leq -0.74\, 10^{-4} $ 
&$-0.74\, 10^{-4}\leq \bar{A}_{CP}\leq 1.71\, 10^{-4} $& I  \\
    \hline
&$0.43\, 10^{-4}\leq \bar{A}_{CP}\leq 3.22\, 10^{-4} $ 
&$-1.05\, 10^{-4}\leq \bar{A}_{CP}\leq 0.43\, 10^{-4} $ & II \\
    \hline \hline

    $0.5$            
&$-1.05\, 10^{-2}\leq \bar{A}_{CP}\leq -1.04\, 10^{-2} $ 
&$-0.25\, 10^{-2}\leq \bar{A}_{CP}\leq 0.55\, 10^{-2} $& I  \\
    \hline
&$0.97\, 10^{-2}\leq \bar{A}_{CP}\leq 1.17\, 10^{-2} $ 
&$-0.34\, 10^{-2}\leq \bar{A}_{CP}\leq 0.14\, 10^{-2} $ & II \\
    \hline \hline

    $0.9$             
&$-3.32\, 10^{-2}\leq \bar{A}_{CP}\leq -0.91\, 10^{-2} $ 
&$-0.90\, 10^{-2}\leq \bar{A}_{CP}\leq 2.72\, 10^{-2} $& I  \\
     \hline
&$0.53\, 10^{-2}\leq \bar{A}_{CP}\leq 2.66\, 10^{-2} $ 
&$-1.86\, 10^{-2}\leq \bar{A}_{CP}\leq 0.53\, 10^{-2} $ & II \\
    \hline \hline
        \end{tabular}
        \end{center} }
\caption{ The average CP asymmetry $\bar{A}_{CP}$ for regions I 
( $1\, GeV\leq \sqrt q^2 \leq m_{J/\psi}-20\, MeV$ ) and II 
($m_{J/\psi}+20\, MeV \leq \sqrt q^2 \leq m_{\psi'}-20\, MeV$ ) }
\label{Table3}
\end{table}

\begin{figure}[htb]
\vskip -3.0truein
\centering
\epsfxsize=6.8in
\leavevmode\epsffile{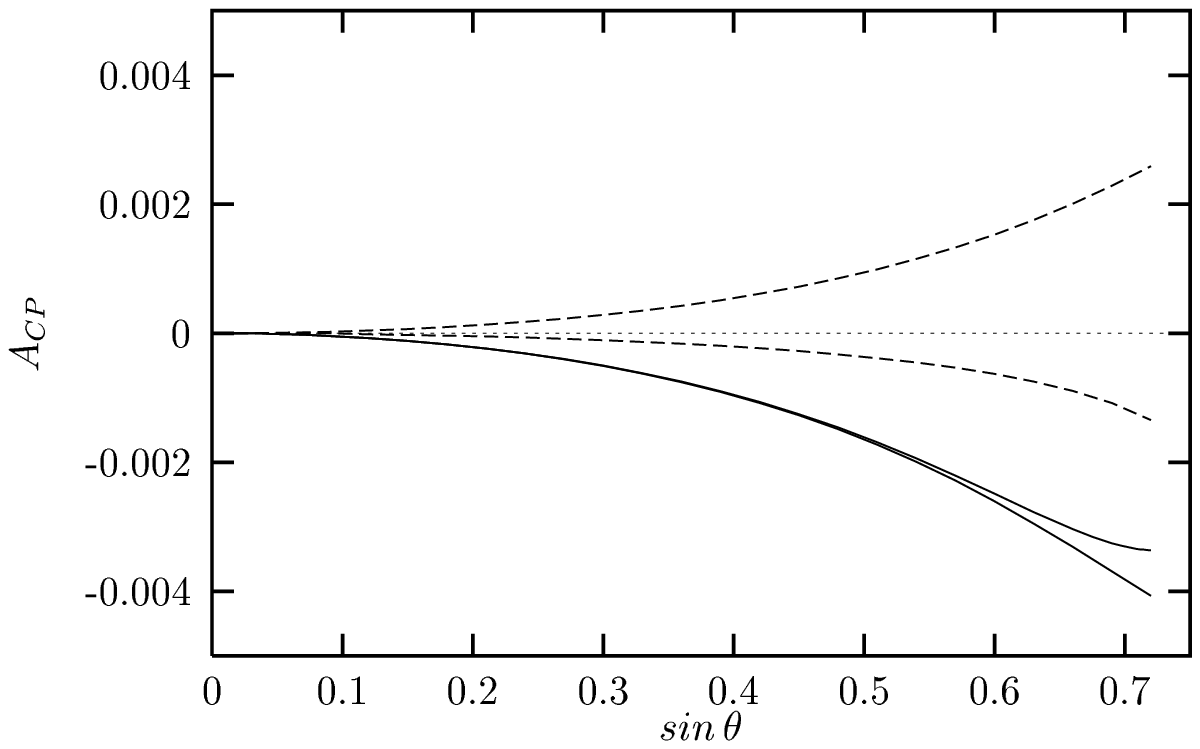}
\vskip -3.0truein
\caption[]{$A_{CP}$ as a function of  $sin\,\theta$ for  
$q^2=4\, GeV^2$, $\bar{\epsilon}_{N,bb}^{D}=40\, m_b$ in the region 
$|r_{tb}|<1$, at the scale $\mu=m_b$, including LD effects. Here 
$A_{CP}$ is restricted in the region bounded by 
solid lines for $C_7^{eff}>0$ and by  dashed lines for $C_7^{eff}<0$.}
\label{cpq24}
\end{figure}

\begin{figure}[htb]
\vskip -3.0truein
\centering
\epsfxsize=6.8in
\leavevmode\epsffile{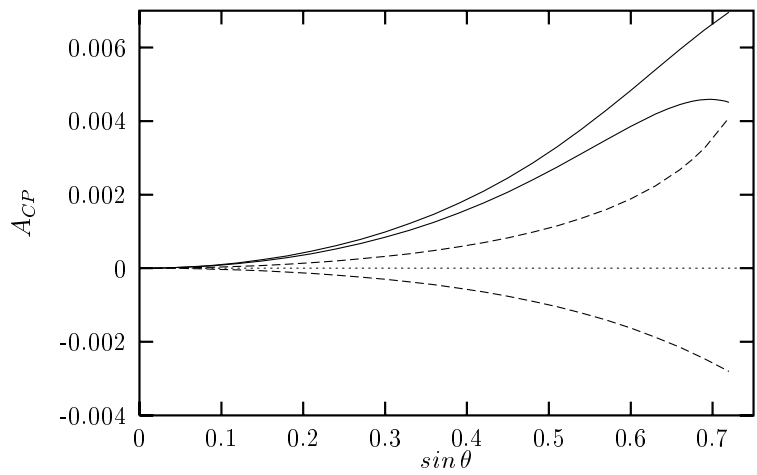}
\vskip -3.0truein
\caption[]{The same as Fig \ref{cpq24}, but for $q^2=12\, GeV^2$.}
\label{cpq212}
\end{figure}

Figs. \ref{cpq24} and \ref{cpq212} are devoted to $sin\,\theta$
dependence of $A_{CP}$ for $q^2=4\, GeV^2$ and 
$q^2=12\, GeV^2$ respectively. Here, $A_{CP}$ lies in the region bounded 
by solid lines for $C_7^{eff} > 0$ or by dashed lines for $C_7^{eff} < 0$.
With decreasing $sin\,\theta$, $A_{CP}$ decreases as expected and  
the restriction region becomes narrower, for both $C_7^{eff} > 0$ and 
$C_7^{eff} < 0$. Further, for fixed $q^2$ values, the sign of $A_{CP}$ does
not change with changing $\sin\,\theta$ for $C_7^{eff} >0$ , contrary to the 
case $C_7^{eff} < 0$. This is informative in the determination of the sign 
of $C_7^{eff}$ with the experimental measurement of $A_{CP}$ at fixed $q^2$. 
Note that the similar situation exist for the general 2HDM with complex 
Yukawa couplings (see \cite{eril2}).

Now we would like to summarize our results:

\begin{itemize}
\item $Br$ for the process under consideration is at the order of $10^{-6}$ 
for $|r_{tb}| <1$ and it is greater for $C_{7}^{eff}>0$ compared to 
$C_{7}^{eff}<0$. Further, it is not sensitive to $sin\,\theta$ especially 
for $C_{7}^{eff}<0$. 

\item $|A_{CP}|$ increases with increasing $sin\,\theta$. 
For $C_{7}^{eff}>0$, $A_{CP}$ changes sign at the $q^2$ value, 
$q^{2}\sim 9 \, GeV^2$, however it can have any sign for $C_{7}^{eff}<0$.
For the case $|r_{tb}|>>1$, $A_{CP}$ almost vanishes ($\sim 10^{-11}$) 
since $sin\,\theta$ should be small due to the restriction coming from the 
limit on neutron electric dipole moment. 

\item  For the fixed value of $q^2$ and $C_{7}^{eff} > 0$, $A_{CP}$ can 
be either negative or positive when $sin\,\theta$ varies. For 
$C_{7}^{eff}<0$, it can have both signs. This shows that with the measurement 
of $A_{CP}$ for fixed $q^{2}$, it is possible to detect the sign of 
$C_{7}^{eff}$, which is an interesting result.

\end{itemize}

Therefore, the experimental investigation of $A_{CP}$ ensure a crucial test
for new physics and also the sign of $C_{7}^{eff}$.
\newpage
{\bf \LARGE {Appendix}} \\
\begin{appendix}
\section{The Wilson coefficients in the SM and the functions appear in
these coefficients}
The initial values of the Wilson coefficients for the relevant process 
in the SM are \cite{Grinstein1}
\begin{eqnarray}
C^{SM}_{1,3,\dots 6,11,12}(m_W)&=&0 \nonumber \, \, , \\
C^{SM}_2(m_W)&=&1 \nonumber \, \, , \\
C_7^{SM}(m_W)&=&\frac{3 x^3-2 x^2}{4(x-1)^4} \ln x+
\frac{-8x^3-5 x^2+7 x}{24 (x-1)^3} \nonumber \, \, , \\
C_8^{SM}(m_W)&=&-\frac{3 x^2}{4(x-1)^4} \ln x+
\frac{-x^3+5 x^2+2 x}{8 (x-1)^3}\nonumber \, \, , \\ 
C_9^{SM}(m_W)&=&-\frac{1}{sin^2\theta_{W}} B(x) +
\frac{1-4 \sin^2 \theta_W}{\sin^2 \theta_W} C(x)-D(x)+\frac{4}{9}, \nonumber \, \, , \\
C_{10}^{SM}(m_W)&=&\frac{1}{\sin^2\theta_W}
(B(x)-C(x))\nonumber \,\, , \\
\label{CoeffW}
\end{eqnarray}
and the primed ones are 
\begin{eqnarray}
C^{\prime SM}_{1,\dots 12}(m_W)&=&0  .
\label{CoeffW2}
\end{eqnarray}
The functions appear in these coefficients are 
\begin{eqnarray}
B(x)&=&\frac{1}{4}\left[\frac{-x}{x-1}+\frac{x}{(x-1)^2} \ln
x\right] \nonumber \,\, , \\
C(x)&=&\frac{x}{4}\left[\frac{x/2-3}{x-1}+\frac{3x/2+1}{(x-1)^2}
       \ln x \right] \nonumber \,\, , \\
D(x)&=&\frac{-19x^3/36+25x^2/36}{(x-1)^3}+
       \frac{-x^4/6+5x^3/3-3x^2+16x/9-4/9}{(x-1)^4}\ln x \,\, ,
\label{BCD}
\end{eqnarray}
and in the coefficients $C_{i}^{(\prime) H}$ (eqs. (\ref{CoeffH}) and 
(\ref{CoeffH2})) are
\begin{eqnarray}
F_{1}(y)&=& \frac{y(7-5y-8y^2)}{72 (y-1)^3}+\frac{y^2 (3y-2)}{12(y-1)^4}
\,\ln y \nonumber  \,\, , \\ 
F_{2}(y)&=& \frac{y(5y-3)}{12 (y-1)^2}+\frac{y(-3y+2)}{6(y-1)^3}\, \ln y 
\nonumber  \,\, ,\\ 
G_{1}(y)&=& \frac{y(-y^2+5y+2)}{24 (y-1)^3}+\frac{-y^2} {4(y-1)^4} \, \ln y
\nonumber  \,\, ,\\ 
G_{2}(y)&=& \frac{y(y-3)}{4 (y-1)^2}+\frac{y} {2(y-1)^3} \, \ln y 
\nonumber\,\, ,\\
H_{1}(y)&=& \frac{1-4 sin^2\theta_W}{sin^2\theta_W}\,\, \frac{x
y}{8}\,\left[ 
\frac{1}{y-1}-\frac{1}{(y-1)^2} \ln y \right]-y \left[\frac{47 y^2-79 y+38}{108
(y-1)^3}-\frac{3 y^3-6 y+4}{18(y-1)^4} \ln y \right] 
\nonumber  \,\, , \\ 
L_{1}(y)&=& \frac{1}{sin^2\theta_W} \,\,\frac{x y}{8}\, \left[-\frac{1}{y-1}+
\frac{1}{(y-1)^2} \ln y \right]
\nonumber  \,\, .\\ 
\label{F1G1}
\end{eqnarray}
\section{3 Higgs doublet model}
We consider three complex, $SU(2)$ doublet scalar fields 
$\phi_i \, (i=1,2,3)$. The gauge and $CP$ invariant Higgs potential which 
spontaneously breaks  $SU(2)\times U(1)$ down to $U(1)$ can be written as:
\begin{eqnarray}
V(\phi_1, \phi_2,\phi_3 )&=&c_1 (\phi_1^+ \phi_1-v^2/2)^2+
c_2 (\phi_2^+ \phi_2)^2 \nonumber \\ &+& 
c_3 (\phi_3^+ \phi_3)^2+
c_4 [(\phi_1^+ \phi_1-v^2/2)+ \phi_2^+ \phi_2+\phi_3^+ \phi_3]^2
\nonumber \\ &+& 
c_5 [(\phi_1^+ \phi_1) (\phi_2^+ \phi_2)-(\phi_1^+ \phi_2)(\phi_2^+ \phi_1)]
\nonumber \\ &+& 
c_6 [(\phi_1^+ \phi_1) (\phi_3^+ \phi_3)-(\phi_1^+ \phi_3)(\phi_3^+ \phi_1)]
\nonumber \\ &+& 
c_7 [(\phi_2^+ \phi_2) (\phi_3^+ \phi_3)-(\phi_2^+ \phi_3)(\phi_3^+ \phi_2)]
\nonumber \\ &+& 
c_8 [Re(\phi_1^+ \phi_2)]^2 +c_9 [Re(\phi_1^+ \phi_3)]^2 +
c_{10} [Re(\phi_2^+ \phi_3)]^2 \nonumber \\ &+&
c_{11} [Im(\phi_1^+ \phi_2)]^2 +c_{12} [Im(\phi_1^+ \phi_3)]^2 +
c_{13} [Im(\phi_2^+ \phi_3)]^2 +c_{14}
\label{potential}
\end{eqnarray}
Here, we assume that only $\phi_1$ has vacuum expectation value 
(see section 2). The parameters $c_i$ are real to ensure the hermiticity 
of the potential term. Further, the Higgs sector does not violate $CP$ and 
all possible $CP$ violation effects are based on the choice of the Yukawa 
couplings.

The Higgs boson squared mass matrix can be obtained by the expression 
\begin{eqnarray}
m_{ij}=\frac{\partial^2  V(\phi_1\,, \phi_2\, ,\phi_3)}
{\partial \theta_i\, \partial\theta_j }\,\, .
\label{massmatr}
\end{eqnarray}
Here $\theta_i$ are real fields satisfying
\begin{eqnarray}
\chi^+&=& \theta_1+ i \theta_2 \nonumber \, , \\
H^+&=& \theta_3+ i \theta_4 \nonumber \, ,\\
F^+&=& \theta_5+ i \theta_6 \nonumber \, ,\\
\chi^0&=& \theta_7 \nonumber \, ,\\
H^2&=& \theta_8 \nonumber \, ,\\
H^4&=& \theta_9 \nonumber \, ,\\
H^0&=& \theta_{10} \nonumber \, ,\\
H^1&=& \theta_{11} \nonumber \, ,\\
H^3&=& \theta_{12}  \, ,
\label{theta}
\end{eqnarray}
where $\chi^+$, $\chi^0$ and $H^0$ are the SM particles and $H^+$, $F^+$, 
$H^i \, (i=1,2,3,4)$ are new particles existing in 3HDM 
(see eq. (\ref{choice}) ). Note that, these fields are mass eigenstates,
thanks to choice eq.(\ref{choice}). Diagonalizing this matrix, we get masses 
of new Higgs particles as:
\begin{eqnarray}
m_{H\pm}&=&\frac{v^2}{2} c_5  \,\, ,\nonumber \\
m_{F\pm}&=&\frac{v^2}{2} c_6  \,\, ,\nonumber \\
m_{H^1}&=&\frac{v^2}{2} c_{8}  \,\, ,\nonumber \\
m_{H^2}&=&\frac{v^2}{2} c_{11}  \,\, ,\nonumber \\
m_{H^3}&=&\frac{v^2}{2} c_{9}   \,\, ,\nonumber \\
m_{H^4}&=&\frac{v^2}{2} c_{12}   \,\, .
\label{masses}
\end{eqnarray}
In eq. (\ref{theta}), $\chi^+$ and $\chi^0$ are golstone bosons, which can be
eaten up in unitary gauge, $H^0$ is the SM Higgs which has mass 
$m_{H^0}=2 v^2 (c_1+c_4)$. $H^1$, $H^3$ are scalar and $H^2$, $H^4$ are
pseudoscalar particles due to new physics. Note that $H^1$ and $H^2$ are
denoted by $h^0$ and $A^0$ in the literature. 

For completeness, we also present the kinetic term for the 3HDM:
\begin{eqnarray}
(D_{\mu} \phi_i)^+ D^{\mu} \phi_i = 
& & (\partial_{\mu} \phi_i^+ + i\frac{g'}{2} B_{\mu} \phi^{+}_i+
i \frac{g}{2} \phi^+_{i} \frac{\vec{\tau}}{2} \vec{W}_{\mu}) 
\nonumber \\
& & (\partial^{\mu} \phi_i - i\frac{g'}{2} B^{\mu} \phi_i-
i \frac{g}{2} \phi_{i} \frac{\vec{\tau}}{2} \vec{W}^{\mu}) 
\label{kinetic}
\end{eqnarray}
where 
\begin{eqnarray}
\phi_{i}=\left(\begin{array}{c c} 
\phi^{+}\\ \phi^0\end{array}\right)  \,\, i=1,2,3\, .
\label{phi}
\end{eqnarray}
\section{The form factors for the decay $B\rightarrow K^* l^+ l^-$ }
The structure functions appear in eq. (\ref{matr2}) are 
\begin{eqnarray}
A &=& -C_9^{eff} \frac{V}{m_B + m_{K^*}} - 4 C_7^{eff} \frac{m_b}{q^2} T_1~,
\nonumber\\ 
B_1 &=& -C_9^{eff} (m_B + m_{K^*}) A_1 - 4 C_7^{eff} \frac{m_b}{q^2} (m_B^2 -
m_{K^*}^2) T_2~,  \nonumber \\ 
B_2 &=& -C_9^{eff} \frac{A_2}{m_B + m_{K^*}} - 4 C_7^{eff} \frac{m_b}{q^2} 
\ga T_2 + \frac{q^2}{m_B^2 - m_{K^*}^2} T_3 \dr~,  \nonumber \\
B_3 &=& -C_9^{eff}\frac{ 2 m_{K^*}}{  q^2}(A_3 - A_0) + 4 C_7 
\frac{m_b}{q^2}T_3~,  \nonumber \\ 
C &=& -C_{10} \frac{V}{m_B + m_{K^*}}~,  \nonumber \\  
D_1 &=& -C_{10} (m_B + m_{K^*}) A_1~,  \nonumber \\     
D_2 &=& -C_{10} \frac{A_2}{m_B + m_{K^*}}~,  \nonumber \\ 
D_3 &=& -C_{10} \frac{2 m_{K^*}}{q^2} (A_3 - A_0)~ .
\label{hadpar1}
\end{eqnarray}
We use the $q^2$ dependent expression which is calculated in the framework 
of light-cone QCD sum rules in \cite{braun} to calculate the hadronic 
formfactors $V,~A_1,~A_2,~A_0,~T_1,~T_2$ and $T_3$: 
\begin{eqnarray}
F(q^2)=\frac{F(0)}{1-a_F \frac{q^2}{m_B^2}+b_F (\frac{q^2}{m_B^2})^2}\, ,
\label{formfac}
\end{eqnarray}
where the values of parameters $F(0)$, $a_F$ and $b_F$ are listed in Table 4.

\begin{table}[h]
    \begin{center}
    \begin{tabular}{|c|c|c|c|}
    \hline
    \hline \hline
                &$F(0)$              &       $a_F$ &  $b_F$\\
    \hline \hline    
    $A_1$       &$0.34\pm 0.05$      &       $0.60$&  $-0.023$ \\
    $A_2$       &$0.28\pm 0.04$      &       $1.18$&  $ 0.281$ \\
    $V  $       &$0.46\pm 0.07$      &       $1.55$&  $ 0.575$ \\
    $T_1$       &$0.19\pm 0.03$      &       $1.59$&  $ 0.615$ \\
    $T_2$       &$0.19\pm 0.03$      &       $0.49$&  $-0.241$ \\
    $T_3$       &$0.13\pm 0.02$      &       $1.20$&  $ 0.098$ \\
    \hline
        \end{tabular}
        \end{center}
\caption{The values of parameters existing in eq.(\ref{formfac}) for 
the various form factors of the transition $B\rightarrow K^*$.} 
\label{Table4}
\end{table}
\end{appendix}

\newpage
\end{document}